\documentclass[preprint2]{aastex}
\begin{document}
\title{Limits on GRB Prompt Radio Emission Using the LWA1}
\author{K.S. Obenberger \\ \affil{Department of Physics and Astronomy, University of New Mexico, Albuquerque NM, 87131}  J.M. Hartman\\ \affil{NASA Jet Propulsion Laboratory, Pasadena, CA 91109 USA \\ NASA Postdoctoral Program Fellow} G.B. Taylor\\  \affil{Department of Physics and Astronomy, University of New Mexico, Albuquerque NM, 87131} J. Craig\\  \affil{Department of Physics and Astronomy, University of New Mexico, Albuquerque NM, 87131} J. Dowell\\  \affil{Department of Physics and Astronomy, University of New Mexico, Albuquerque NM, 87131} J.F. Helmboldt\\  \affil{US Naval Research Laboratory, Code 7213, Washington, DC 20375} P.A. Henning\\  \affil{Department of Physics and Astronomy, University of New Mexico, Albuquerque NM, 87131} F.K. Schinzel\\  \affil{Department of Physics and Astronomy, University of New Mexico, Albuquerque NM, 87131} T.L. Wilson\\  \affil{US Naval Research Laboratory, Code 7213, Washington, DC 20375}}

\begin{abstract}
As a backend to the first station of the Long Wavelength Array (LWA1) the Prototype All Sky Imager (PASI) has been imaging the sky $>$ -26$^{\circ}$ declination during 34 Gamma Ray Bursts (GRBs) between January 2012 and May 2013. Using this data we were able to put the most stringent limits to date on prompt low frequency emission from GRBs. While our limits depend on the zenith angle of the observed GRB, we estimate a 1$\sigma$ RMS sensitivity of 68, 65 and 70 Jy for 5 second integrations at 37.9, 52.0, and 74.0 MHz at zenith. These limits are relevant for pulses $\geq$ 5 s and are limited by dispersion smearing. For pulses of length 5 s we are limited to dispersion measures ($DM$s) $\leq$ 220, 570, and 1,600 pc cm$^{-3}$ for the frequencies above. For pulses lasting longer than 5s, the $DM$ limits increase linearly with the duration of the pulse. We also report two interesting transients, which are, as of yet, of unknown origin, and are not coincident with any known GRBs. For general transients, we give rate density limits of $\leq$ $7.5\times10^{-3}$, $2.9\times10^{-2}$, and $1.4\times10^{-2}$  yr$^{-1}$ deg$^{-2}$ with pulse energy densities $>1.3\times 10^{-22}$,  $1.1\times 10^{-22}$, and $1.4\times 10^{-22}$ J m$^{-2}$ Hz$^{-1}$ and pulse widths of 5 s at the frequencies given above.\\
\end{abstract}

\section{Introduction}

Since the discovery of Gamma Ray Bursts (GRBs) by Klebesadel et al. (1973) there have been several groups to propose mechanisms capable of producing prompt low frequency ($< $100 MHz) radio emission observable from Earth. Usov \& Katz (2000) suggested that low frequency radiation could be created by oscillations in the current sheath that separates a strongly magnetized jet and the surrounding ambient plasma. This emission would peak at 1 MHz and drop off following a power law at higher frequencies. The bulk of the emission lies below the ionospheric cutoff of about 10 MHz, but the high frequency tail of this might extend up to frequencies observable by ground based telescopes. The flux density of the high frequency tail is approximated with a power law $\propto \nu^{-1.6}$. As an example they provide a best case estimate of $\sim$10$^{2}$ Jy at 30 MHz.


Sagiv \& Waxman (2002) also predict low frequency emission to occur in the early stages of the afterglow (10s after the GRB). In this scenario a strong synchrotron maser condition is created at frequencies below 200 MHz, due to an excess of low energy electrons. The excess is created by a build up of injected electrons that cool to low energies through synchrotron radiation. The effect is amplified when the jet propagates into a medium denser than the ISM. Such a dense environment would exist around high mass Wolf-Rayet stars, which are thought to be the progenitors of long duration GRBs. 

While no prompt low frequency emission has yet been detected, a future detection would yield a number of constraints on the parameters of GRBs. The dispersion measure (DM) of prompt radio emission would allow estimates of the physical conditions of the region immediately surrounding nearby ($z$ $\lesssim$ 0.5)\footnote{A redshift of 0.5 is chosen because above this point the DM contribution from the intergalactic medium would be roughly equal to the maximum contribution from a galaxy similar to our own (Ioka, 2003). However if the DM of the host galaxy is larger than that of our own, ``nearby'' would include larger redshifts.} GRBs, telling us about the environment in which GRB progenitors are formed. For more distant GRBs the DM would be dominated by the InterGalactic Medium (IGM), thus giving a measurement of the number of baryons in the universe (Ginzburg 1973). For extremely distant ($z$ $>$ 6) GRBs a dispersion measure could act as a probe of the reionization history (Ioka 2003).

Over the past three decades there have been many searches for prompt, low frequency GRB emission (Baird et al. 1975, Dessenne et al. 1996, Koranyi et al. 1995, Benz \& Paesold, 1998, Balsano 1999, Morales et al. 2005, Bannister et al. 2012). Of these studies, 2 have been below 100 MHz. Benz \& Paesold (1998) covered the range from 40 - 1000 MHz, had a RMS sensitivity of $\sim 10^{5}$ Jy, and observed during 7 GRBs between February of 1992 and March of 1994. Balsano (1999) covered 72.8 - 74.7 MHz, observed 32 GRBs between September of 1997 and March of 1998, and had a wide range in root mean square (RMS) sensitivities for each GRB. The best limit reported in Balsano (1999) was $\sim$200 Jy for 50 ms integrations. Both of these studies used BATSE triggers, which had a position uncertainty typically around a few degrees. Morales et al. (2005) reported on a planned study centered at 30 MHz.

In this paper we present a search for prompt low frequency emission from 34 GRBs using the all-sky imaging capabilities of the Prototype All Sky Imager (PASI), a backend to the first station of the Long Wavelength Array (LWA1). While our objective was to find or place limits on prompt emission from GRBs, we also conducted a search for generic transients occurring during our observations but located elsewhere in the sky. In \S2 we describe the LWA1 telescope and the PASI backend, in \S3 we describe the NASA GCN/TAN (Gamma- ray Coordinates Network / Transient Astronomy Network) and how we made use of it for our observations, in \S4 we discuss how dispersion would effect prompt emission from GRBs, in \S5 we discuss our data, analysis and our results, in \S6 we describe our search for generic transients and our results, in \S7 we describe rate density limits on generic transients, and \S8 is a discussion of our findings.

\section{The Long Wavelength Array Prototype All Sky Imager}
Co-located with the Karl Jansky Very Large Array (VLA), LWA1 is one of several low frequency radio telescopes currently coming online that are searching for transients. Among others are LOFAR (Low Frequency Array, van Haarlem et al 2013) and MWA (Murchison Widefield Array, Bowman et al. 2013).

 LWA1(Ellingson et al. 2013; Taylor et al. 2012) consists of 256 cross-dipole antennas (512 dipoles) spread over an ellipse of 100x110m elongated in the N-S direction. Surrounding the filled aperture of LWA1 are 5 individual cross-dipoles at distances of about 200 - 500 m distance from the center of the array. While LWA1 is operating in the the narrow band transient buffer (TBN) mode, 100 kHz of bandwidth (75 kHz usable) is continuously read out and is tunable to any center frequency between 10 - 88 MHz.

Near realtime imaging is done by the Prototype All Sky Imager (PASI), a small dedicated computing cluster, which correlates and images a live stream of TBN. PASI produces a set of visibilities every 5 seconds, and since each antenna sees the entire sky, the images produced are of the entire sky.  The 5 second visibilities are stored as CASA measurement sets on a 4 TB external hard drive buffer. The lifetime of a 5 second visibility set on the hard drive is about 2 weeks. PASI also images the visibilities and displays them on the LWA TV\footnote{http://www.phys.unm.edu/$\sim$lwa/lwatv.html} website and stores time-lapse movies from each day (Obenberger et al., in prep).

While PASI has recorded at dozens of different frequencies since it has been operating, the majority of the time it has been at 37.9, 52.0, and 74.0 MHz. Being within protected radio astronomy bands 37.9 and 74.0 MHz are subjected to a minimal amount of RFI. While 52.0 MHz is not protected, the amount of RFI is typically similar. For this paper we have respectively analyzed 112.6, 29.7, and 59.8 hours at the above frequencies. These hours represent both the data taken during the GRBs and the data taken for calibration purposes.

Using the CASA clean algorithm (McMullin et al. 2007), PASI produces dirty images that have not been flux calibrated, self-calibrated, flagged for RFI, or deconvolved. An example all sky image is shown in figure 1.

\begin{figure}
	\centering
	\includegraphics[width = 3in]{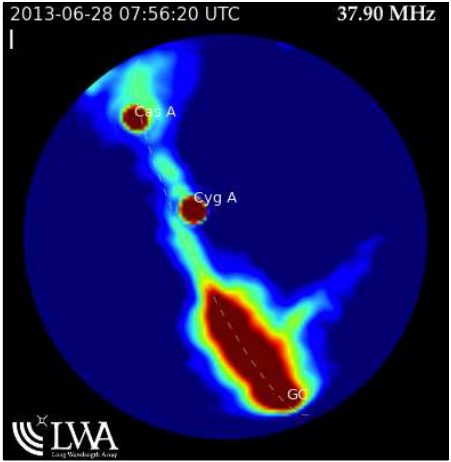}
	\caption{Example all-sky image produced by PASI with a dynamic range of $\sim$ 50 dB. While we only use zenith angles less than 60$^{\circ}$ this image shows the full $2 \pi$ sr. The Galactic plane, Galacic Center (GC), Cygnus A (Cyg A), and Cassiopeia A (Cas A), can all be seen in the image.}
\end{figure}

For this paper flux calibration at 74.0 MHz was achieved by fitting the measured flux of Taurus A with its known flux density of 1811.3 $\pm$ 3.07 Jy at 74 MHz from the VLA Low-Frequency Sky Survey (VLSS) provided in Kassim et al. (2007). The same was done at 52.0 and 37.9 MHz using scaled flux densities from Baars et al. (1977). Fitting a third order polynomial to the measurement of Taurus A as it transits the sky, we were able to model the zenith angle dependent power pattern to a good approximation. Figure 2 shows the normalized measured power pattern of Taurus A.

Our RMS sensitivity is a function of zenith angle and frequency. To estimate these dependencies we measured the pixel noise for a quiet off source region as it transits the sky, passing through zenith. We then scaled the amount of scatter with the power pattern derived from Taurus A. Finally we calculated the RMS of the scaled noise every 50 integrations (250 s or 1$^{\circ}$ of change). For each frequency, we repeated this process on 2 separate days each day having a different off source location. This resulted in averaged zenith angle RMS sensitivity estimates, with zenith values of 68, 65 and 70 Jy for 37.9, 52.0, and 74.0 MHz for 5 second integrations. Figure 3 shows a scatter plot of the RMS for 37.9 MHz calculated using the method described above on the 2 occasions mentioned above. A lognormal model yielded a good fit for zenith angles $<$ 65$^{\circ}$.

\begin{figure}
	\centering
	\includegraphics[width = 3in]{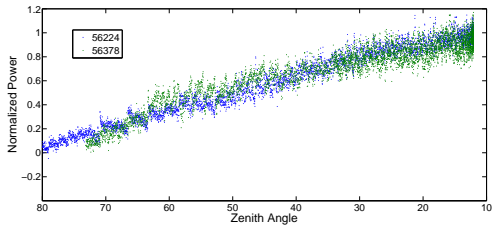}
	\caption{The normalized measured power pattern of Taurus A as a function of zenith angle, measured on two separate days. Dates in MJD. The large fluctuations at greater zenith angles are most likely caused by the ionosphere.}
\end{figure}

\begin{figure}
	\centering
	\includegraphics[width = 3in]{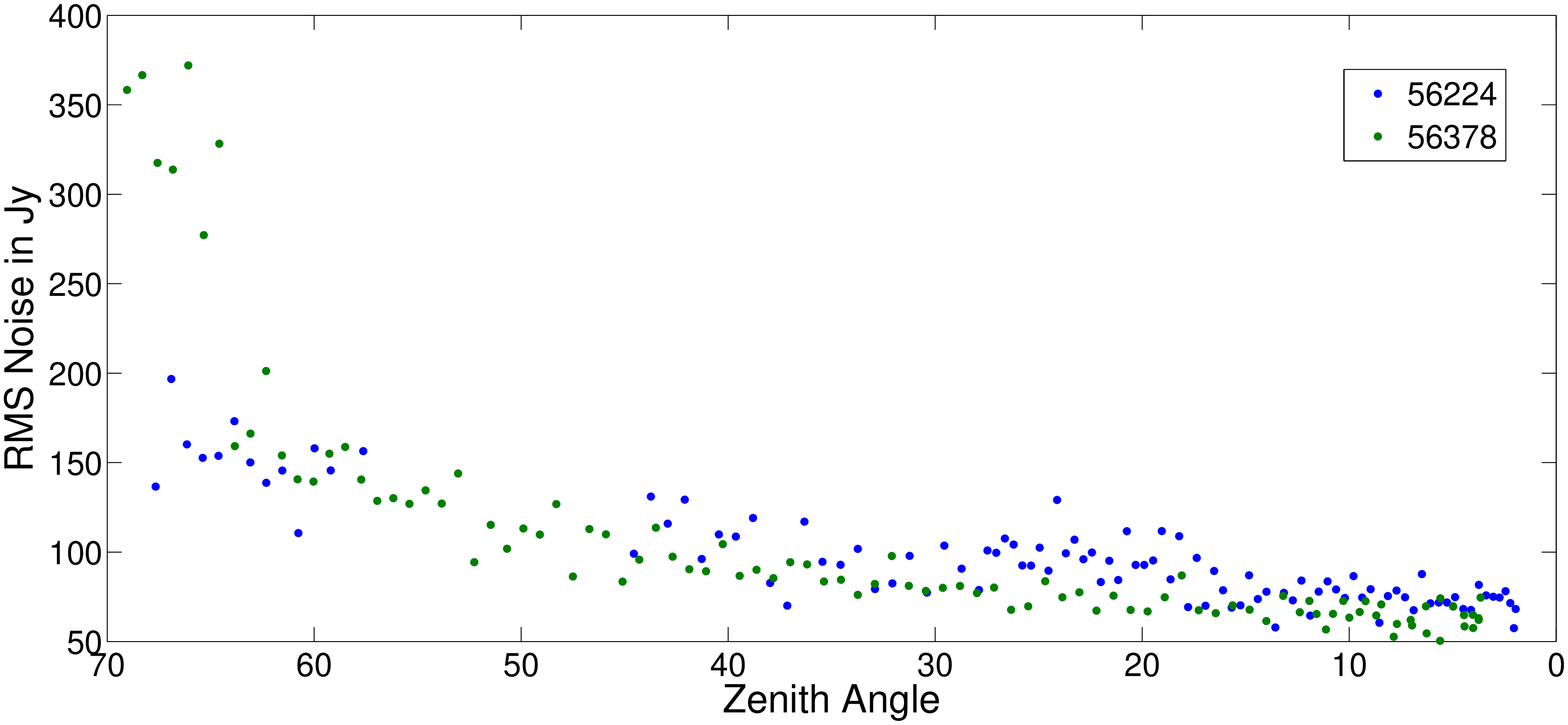}
	\caption{Estimated zenith angle dependent RMS sensitivity for 37.9 MHz, calculated for two separate days. Dates in MJD.}
\end{figure}

The spectrum of each visibility is broken into 6 channels, covering the 75 kHz of bandwidth. While this is not a large bandwidth, it is sufficient to exclude much of the Radio Frequency Interference (RFI), which is often narrower than our 75 kHz. RFI is also excluded based on the location on the sky; transients with zenith angle greater than 60$^\circ$ are not considered to be of celestial origin. PASI also preserves all four Stokes parameters, which can be used for further analysis of transient events.

\section{GCN Trigger Archive}

Currently we use the NASA GCN/TAN (Gamma-ray Coordinates Network / Transient Astronomy Network)  trigger archive\footnote{http://gcn.gsfc.nasa.gov} (Barthelmy et al. 1995) from Swift (Burrows et al. 2005; Barthelmy et al. 2005), Fermi GBM (Gamma-ray Burst Monitor) (Meegan et al. 2009; Briggs et al. 2009), and MAXI (Monitor of All-sky X-ray Image) (Matsuoka et al., 2009). MAXI provides the fewest triggers with about 1 GRB per month, and has 1 arcmin resolution. Swift is more favorable with 2 GRBs per week and arcsec resolution with the XRT (X-Ray Telescope) and UVOT (UV/Optical Telescope)  and arcmin with BAT (Burst Alert Telescope) covering the electromagnetic spectrum from optical to hard X-rays. Fermi GBM is the most prolific with a large field of view enabling the detection of 20 GRBs per month, but only a position accuracy of 1-10 degrees at best. 

For this study we have used the GCN GRB triggers that occurred during PASI operation between January 1 2012 and May 25 2013. Furthermore we excluded triggers with initial zenith angles greater than 60$^{\circ}$ (with 6 exceptions\footnote{To account for possible unknown, intrinsic delays, these 6 triggers were selected to provide a sample that would be at smaller zenith angles several hours after the $\gamma$-ray emission arrived.} in the eastern sky) and those that occurred during periods of exceptionally high RFI.

\section{Delay and Dispersion}

A prompt pulse is delayed and stretched out in time at low frequencies. A dispersion measure, $DM$, is proportional to the number of electrons between an observer and a source. For a pulse of constant frequency (expansion is negligible) the delay time of a pulse of frequency $\nu$ with $DM$ is given by:

\begin{equation}
\tau(\nu) = k_{DM} \times DM \times \frac{1}{\nu^{2}}
\end{equation}

Where the dispersion constant $k_{DM}$ = $ e^{2} / (2 \pi m_{e} c) $ = 4149 MHz$^{2}$ pc$^{-1}$ cm$^{3}$ s. The total dispersion measure can be broken up into three regions, our own galaxy DM$_{Gal}$ the inter galactic medium DM$_{IGM}$, and the GRB's host galaxy DM$_{host}$. The DM of our galaxy varies based on the orientation to the plane, and ranges from DM$^{min}_{Gal} \sim 30$ pc cm$^{-3}$ to DM$^{max}_{Gal} \sim 10^{3}$ pc cm$^{-3}$ (Taylor et al. 1993; Nordgren et al. 1992). Using equation 1 this range of DM correspond to a range in time delay of  46 s to 25 min at 52 MHz.

The DM$_{IGM}$ depends on the number of electrons in the intergalactic medium between us and the GRB and is therefore dependent on the redshift of the galaxy. However, because of the expansion of the universe, the frequency of the radiation emitted in our direction will decrease. Thus a generalization of Eq. 1 must be used to calculate the delay caused by DM$_{IGM}$. Ioka 2003 estimates the redshift dependent DM$_{IGM}$ would be:

\begin{figure}[!]
	\centering
	\includegraphics[width = 3in]{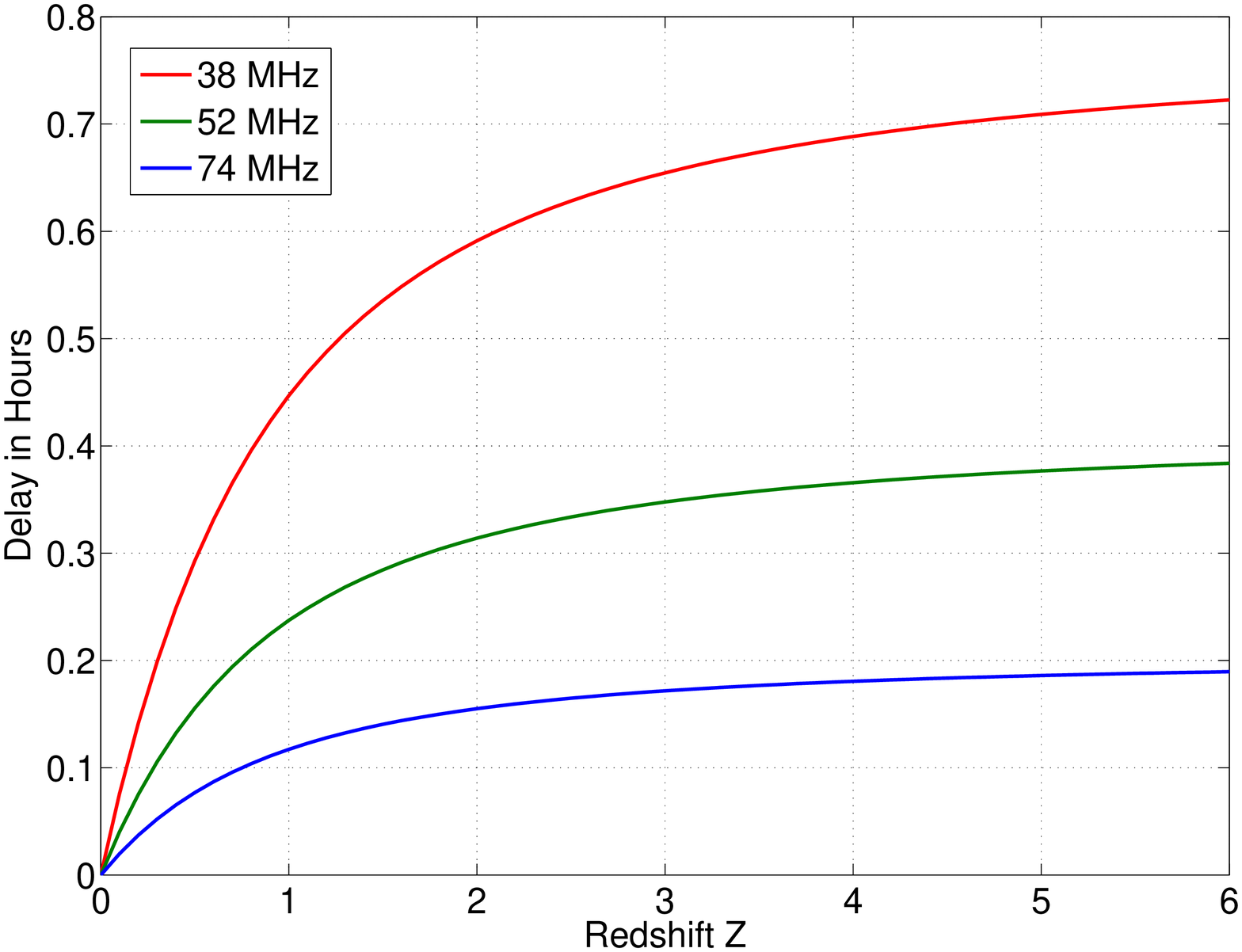}
	\caption{Time delay due to the IGM dispersion for 37.9, 52.0 and 74.0 MHz, as a function of redshift. These are numerical calculations of equation 4 derived from equation 2 of Ioka 2003. }
\end{figure}

\begin{equation}
DM_{IGM} = \frac{3 c H_{0} \Omega_{b}}{8 \pi G m_{p}} \int_0^z \frac{(1+z)dz}{[\Omega_{m}(1+z)^{3} +\Omega_{\Lambda}]^{1/2}}
\end{equation}

The actual frequency, $\nu$, of a photon seen by the electrons at redshift z is related to the observed frequency, $\nu_{ob}$ by:

\begin{equation}
\nu = \nu_{ob}\times(1+z)
\end{equation}

Therefore the time delay of an emitted photon observed to be at frequency, $\nu_{ob}$, is given by:
\begin{equation}
 \tau(\nu) = \frac{3 c H_{0} \Omega_{b}}{\nu^{2}_{ob}8 \pi G m_{p}} \int_0^z \frac{dz}{(1+z)[\Omega_{m}(1+z)^{3} +\Omega_{\Lambda}]^{1/2}}
\end{equation}

With $(\Omega_{m},\Omega_{\Lambda},\Omega_{b},h)$ = (0.3175, 0.6825, 0.04810, 0.6711), (Plank Collaboration, 2013) ,the values for the time delay, $\tau(\nu_{ob} )$, due to the intergalactic medium, are shown in figure 4. $DM_{host}$ is the most difficult quantity to estimate, and may dominate the total DM in some cases. Examining at equations 1 and 3 it is easy to see that at higher redshifts the DM of a host galaxy similar to ours would play a very small role in the delay. However if the GRB progenitors dwell in dense star forming regions, such as the case may be for some long GRBs, then a very small optical depth may make it difficult to detect any signal at all.

Since we do not dedisperse the bursts, we are subject to dispersion smearing across our entire 75 kHz band. In seconds the dispersion smearing time is given by:

\begin{equation}
\tau_{smear} = 8242 \times DM  \times \nu^{-3} \times \Delta \nu
\end{equation}

Where $\nu$ is the center frequency in MHz and $\Delta \nu$ is the bandwidth in MHz. Therefore bursts $\leq$ 5 seconds would become smeared out across our band to at least 1.5 times their duration at $DM$s  $\geq$ 220, 570, and 1,600 pc cm$^{-3}$ at 37.9, 52.0, and 74.0 MHz. However, if the bursts are of longer duration, then dispersion smearing has less of an effect on the sensitivity to those bursts, and we can see out to larger $DM$s. For instance a burst of 30 seconds would be smeared out to 1.5 times their duration at $DM$s $=$ 1,300, 3,400, and 9,800 pc cm$^{-3}$ at those frequencies.

\section{Data, Analysis, and Results}
When given a set of GRB coordinates, a date, and time, we used our archived visibility data to image the entire sky and track the GRB's position. While we expect a maximum combined delay of $<$ 1 hr from the IGM and the Milky Way we allow for large delays from the host galaxies. We analyze the 2 hours after the prompt $\gamma$-ray emission for observations at 74.0 MHz, 3 hours for 52.0 MHz and 4 hours for 38.0 MHz. Tracking of the source is accomplished by mapping the images onto an right ascension (RA) and declination (DEC) coordinate system and selecting the pixels of one beam around the source location. We can then analyze the light curve of that region and look for detections.

From our data, there have not been any significant ($> 5\sigma$) transient events corresponding spatially and temporally to any GRB triggers. However we can provide limits to the peak intensity of a transient.

Given the zenith angle of the GRB at the time the $\gamma$-ray emission arrives, we estimated the 1 $\sigma$ limit with the RMS sensitivity model described in section 2. See Table 1 for our list of GRBs and estimated limits\footnote{While there were $\sim$500 GRBs detected by GCN/TAN telescopes during the 17 months of our study, PASI only observes 1/4 of the total sphere. Moreover PASI was only fully operating during about 1/3 of this time. Also there were a handful of events which occurred during periods of high RFI or solar activity, which prevented us from making reliable limits. This is why we only have 34 GRB observations.}. Since we are using only the dirty images from PASI, which undergo neither a phase self calibration nor deconvolution, these estimates can be improved following planned algorithm development work.

These limits reflect the RMS noise for 5 second integrations. If a pulse was shorter than 10 seconds, our signal to noise (S/N) would be decreased by the ratio of time spent in any one time bin to the size of the bin itself. For instance consider a 7 second, 1000 Jy pulse, that occurred at zenith at 38 MHz and spent 3 seconds in the first bin and 4 seconds in the second. In the second bin the observed flux density would be $1000 \times (0.8)$ Jy. Therefore our S/N would decrease by a factor of $0.8$.

\begin{deluxetable}{ |c| |c| |c| |c| |c|  }
\tablecaption{GRBs and their limits.}
\tablewidth{0pt}
\tablehead{
\colhead{Telescope} 	&	\colhead{Trigger \#}	&	\colhead{Name}	&	\colhead{Frequency}		&	\colhead{1$\sigma$ RMS} \\ &&& (MHz)& (Jy)}
\startdata
Fermi 		&	347464837	&	120105584		&	74.0		&	82 \\
Fermi		&	347831163	&	120109824		&	74.0		&	103 \\
Fermi		&	348250807	&	120114681		&	74.0		&	121 \\ 
Swift			&	519211		&	120403A			&	74.0		&	121 \\
Fermi		&	356223561	&	120415958		&	74.0		&	126 \\
Fermi 		&	356646915	&	120420858		&	74.0		&	139 \\
Fermi 		&	357141744	&	120426585		&	74.0		&	135 \\
Fermi		&	357182249	&	120427054		&	74.0		&	176 \\
Fermi		&	358605842	&	120513531		&	74.0		&	155 \\
Fermi		&	359894162	&	120528442		&	74.0		&	99 \\
Fermi		&	360039223	&	120530121		&	74.0		&	117 \\
Fermi		&	360586337	&	120605453		&	74.0		&	128 \\
\hline
Fermi		&	364049062	&	120715066		&	52.0		&	112 \\
Fermi             	&	364139465	&	120716577		&	52.0		&	133 \\
Fermi 		&	364151106	&	120716712		&	52.0		&	158 \\
Swift	 		&	529095		&	120729A			&	52.0		&	79 \\ 
Fermi		&	367031309	&	120819048		&	52.0		&	 116\\
\hline
Fermi		&	372317712	&	121019233		&	37.9		&	 107\\
Fermi		&	374504566	&	121113544		&	37.9		&	 194\\
Fermi		&	374804740	&	121117018		&	37.9		&	 97\\
Fermi		&	375534890	&	121125469		&	37.9		&	 112\\
Swift	 		&	539866		&	121128A			&	37.9		&	 153 \\ 
Swift	 		&	540964		&	121209A			&	37.9		&	 123 \\
Swift	 		&	544784		&	130102A			&	37.9		&	 104 \\
Fermi		&	379209148	&	130106995		&	37.9		&	 137\\
Fermi		&	381843217	&	130206482		&	37.9		&	 125\\
Swift	 		&	548760		&	130215A			&	37.9		&	 93 \\ 
Fermi		&	383388785	&	130224370		&	37.9		&	 99\\
Swift	 		&	552063		&	130327A			&	37.9		&	 90\\
Fermi		&	387054766	&	130407800		&	37.9		&	 88\\
Swift	 		&	553918		&	130419A			&	37.9		&	 130\\
\tablenotemark{7}Swift	 		&	554620		&	130427A			&	37.9		&	 117\\
MAXI	 	& 	418849999	&	N/A				&	37.9		&	 103\\
Swift	 		&	556344		&	130521A			&	37.9		&	 100\\
\enddata
\tablenotetext{7}{For Swift GRB 120427A, PASI was not operating for the first 30 minutes after the GRB.}

\end{deluxetable}

\section{Interesting Transients Not Associated with GRBs}

Throughout the course of searching for prompt radio emission from GRBs we also searched our data for generic transients in our field of view. We automated our search using image subtraction methods, which inherently increased our noise by $\sqrt{2}$ but allowed us to investigate changes on the order of 10 s by subtracting the third previous image from every image and setting the threshold at 6$\sigma$. Below 6$\sigma$ the number of events detected displayed gaussian behavior as expected with an additional bump at 5 $\sigma$ from false detections from RFI. Above 6$\sigma$ we found 18 events at 37.9 MHz, 7 at 52.0 MHz, and 2 at 74.0 MHz. Except for 1 event at 37.9 MHz and 2 events at 52.0 MHz, nearly all of these events were immediately identified as RFI.

The two events at 52.0 MHz each lasted for only one integration (5s) and appear to be broadband across the 75 kHz. However, upon further investigation these events were found to have significant linear polarization, which indicates that they were most likely reflected man-made RFI, possibly from the ionosphere or from ionized trails left by meteors.

The 37.9 MHz event, however, was a Fast Rise / Exponential Decay (FRED) transient candidate. The event occurred on 2012 Oct. 24 (121024) at  08:37:39 UT, lasted for 75 seconds, had an RA and DEC of 04h 14m 00s +76d 54m 00s with an estimated error of $\sim 1.5^{\circ}$. The light curve of the transient displayed a rise time of $\sim$ 15 s and decay of $\sim$ 60 s (Fig. 5). At peak intensity, this source appears to be $\sim$ 2.4 kJy, and is constant across the 75 kHz band. However upon examining all 4 stokes parameters, there is a slight bump of linear (-U) and left hand circular (-V) as shown in figure 5. The exact percentage of polarization is difficult to quantify since both the -U and -V components each last for one integration and are both $<5\sigma$. When compared to the entire 75 s burst the -U and -V components are $5 \pm 1\%$ and $4 \pm 1\%$. While this polarization may be real, instrumental leakage is very likely the cause. The leakage into the three stokes polarizations on the LWA1 is a function of a sources position on the sky but has not yet been characterized. As a reference we measured the leakage of Cassiopeia A, an unpolarized source at 38 MHz, during the same period the transient was detected. During this time we measured -Q, U, and V leakages of $\sim$ 3$\%$, 8$\%$, and 3$\%$. Cassiopeia A was at approximately the same zenith angle as the transient.

A second event occurred on 2012 Nov. 18 (121118) at 09:53:40 UT, lasted for 100 seconds, had an RA and DEC of 07h 22m 24s +41d 18m 00s, and was observed at 29.9 MHz. The light curve shows similar properties to the 121024 event with a rise time of $\sim$ 25 s and decay time of $\sim$75 s, a maximum flux density of 3.2 kJy, and is also constant across the band. However during this event there were no detectable polarized components (Fig. 5).

Examining the 75 kHz bandwidths we see no signs of any dispersion for either event. However we are able to limit the $DM$s of the 121024 and 121118 events to be approximately $\leq$ 450 and $\leq$  250 pc cm$^{-3}$.

\begin{figure}[!]
	\centering
	\includegraphics[width = 3in]{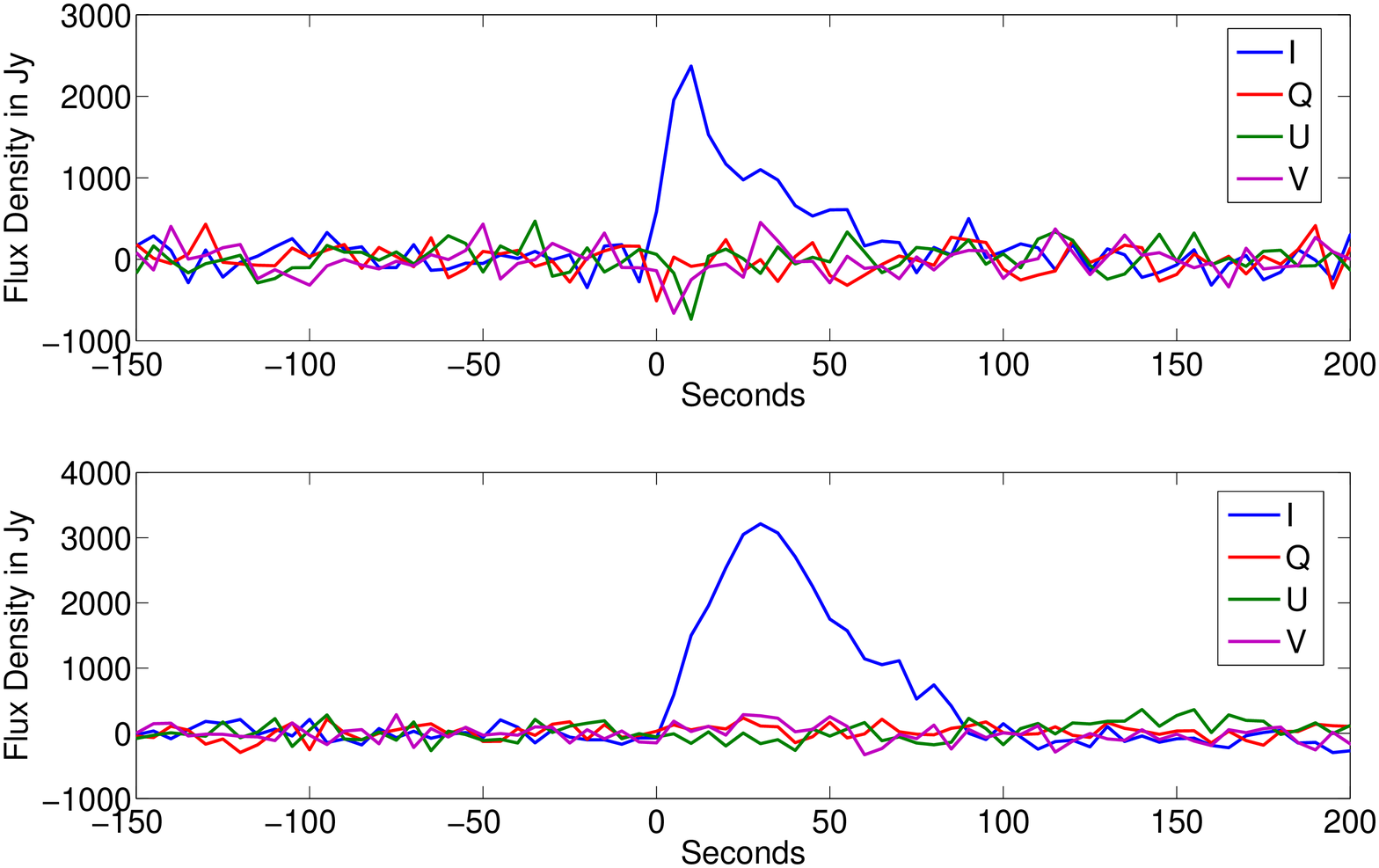}
	\caption{(Top) Light curve of the 121024 event, at 37.9 MHz. The -U burst is at 10 s and the -V burst is at 5 s. (Bottom) Light curve of the 121118 event, at 29.9 MHz.}
\end{figure}

Using the NASA/IPAC Extragalactic Database\footnote{http://ned.ipac.caltech.edu} we found 5 sources above 20 Jy at 38 MHz within $3^{\circ}$, twice our estimated position error, of the 121024 event and none above 20 Jy at 38 MHz within $3^{\circ}$ of the 121118 event. All of the 121024 sources were part of the revised source list of the Rees 38-MHz (8C) survey (Hales et al. 1995; Rees 1999), and no additional sources were found using that catalog. Also there were no additional sources above 7 Jy at 74 MHz found using the VLSS within $3^{\circ}$ of either event (Cohen et al. 2007). 
\begin{figure}[!]
	\centering
	\includegraphics[width = 3in]{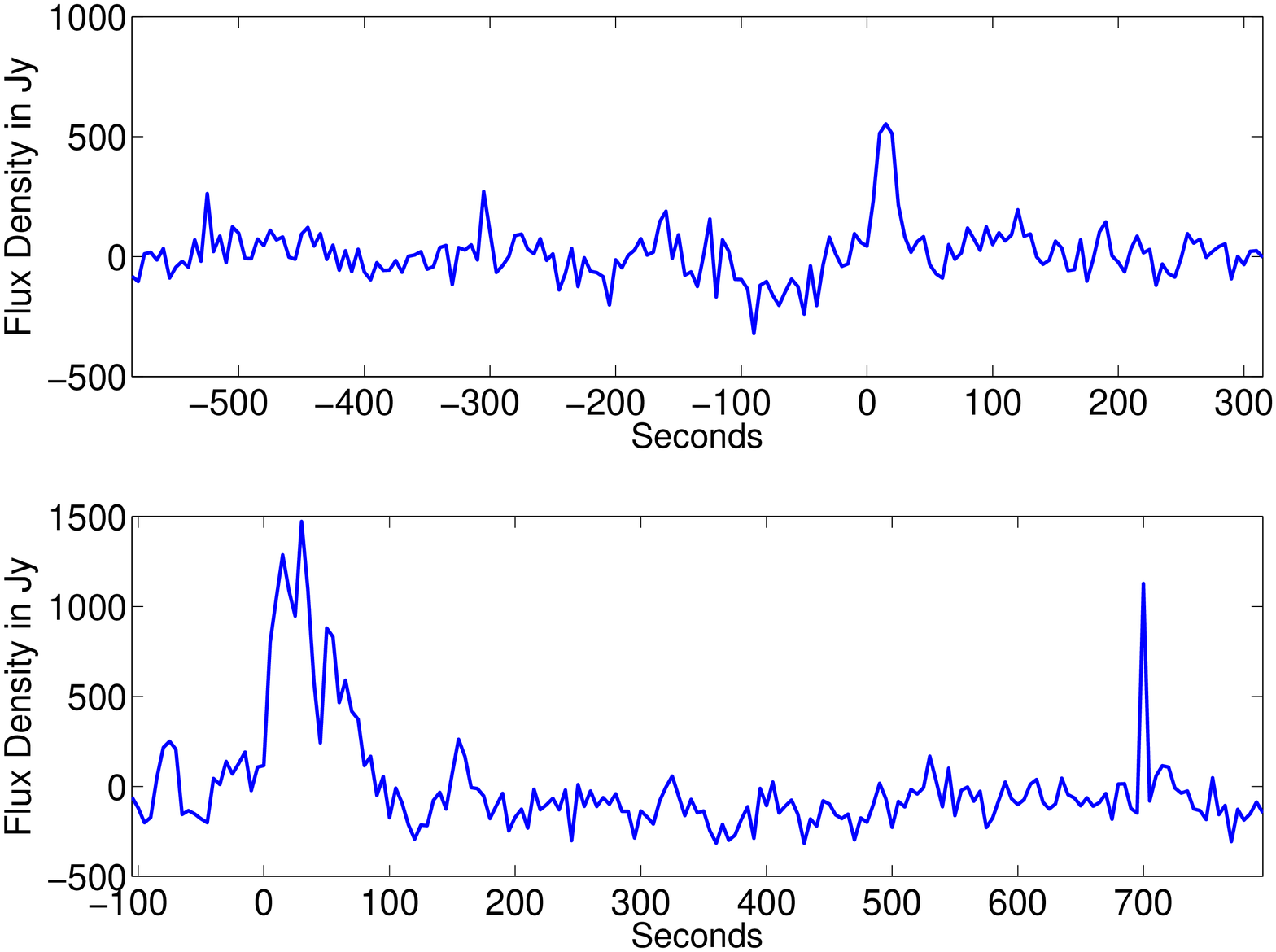}
	\caption{Light curves 3C249 (Top) and 3C230 (Bottom). Each source has been magnified far above their actual flux densities (37 and 76 Jy) due to ionospheric focusing. 3C230 shows as second brightening lasting for only one 5 second integration 700 seconds after the initial brightening. }
\end{figure}

Table 2 lists the sources near the 121024 event. It is possible for any one of these 5 sources to be focused by the ionosphere and temporarily increase in brightness. In fact this is a regular occurrence with sources in our field of view. There will often be periods when several sources on the sky will fluctuate up to 15 times their normal brightness. The effect usually covers the entire sky, in that many sources across the sky will fluctuate (shimmer) for up to several hours. This is likely caused by turbulence in the ionosphere; similar variation has been observed by other instruments at the same frequency (Bezrodny et al. 2008).

At these times of high shimmering, sources that lie below our detectable limit will sometimes be magnified above our threshold and appear for a short period of time. While the vast majority of events occur for sources above 100 Jy, there have been 4 at lower flux densities in the 112 hours of observations at 37.9 MHz data reported in this paper. The typical shape for a light curve of one of these events is a fast rise, fast decay, often lasting for just one integration. Occasionally the source will stay bright for up to a minute, displaying several peaks as it dims and brightens. Figure 6 shows two typical light curves from brightening events. The first is 3C249, which is the dimmest object (37 Jy at 38 MHz) to be magnified above 6$\sigma$ in the data reported in this paper. The second is 3C230 (76 Jy at 38 MHz), which displayed a brightening event lasting $\sim$ 75s, during which it peaked several times.

\begin{deluxetable}{ |c| |c| |c| |c| |c| }

\tablecaption{Sources above 20 Jy within $3^{\circ}$ of the 121024 event}
\tablewidth{0pt}
\tablehead{\colhead{Name} 		&	\colhead{RA}		&	\colhead{DEC}	&	\colhead{S$_{int}$ (Jy)}		&	\colhead{Dist}}

\startdata
8C 0422+770		&	04h 29m 19s		&	+77d 09m 13s	&	42.8				&	0.9$^{\circ}$\\
8C 0357+747		&	04h 03m 15s		&	+74d 55m 58s	&	35.1				&	2.1$^{\circ}$\\
8C 0407+747		&	04h 13m 16s		&	+74d 51m 05s	&	28.5				&	2.0$^{\circ}$\\
8C 0343+749		&	03h 49m 52s		&	+75d 09m 01s	&	22.0				&	2.3$^{\circ}$\\
8C 0415+763		&	04h 22m 06s		&	+76d 27m 05s	&	21.8				&	0.6$^{\circ}$\\
\enddata

\end{deluxetable}

There are several reasons why we believe the 121024 event was not simply one of these focusing events. The first is that this would be one of the strongest focusing events we have seen that is a factor of $\sim$ 60 if it is 8C 0422+770 and $\sim$ 120 if it is 8C 0415+763. The second is that during the hours before and after, the other sources in the sky were shimmering only slightly. Finally the light curve is very similar to the light curve of the 121118 event for which there is no corresponding bright sources and is dissimilar to a light curve of a typical shimmering event. Therefore it is our belief that these events are not ionospheric focusing of objects just below our sensitivity limit. 

Many astrophysical sources have been theorized to produce low frequency transient emission. Possible sources include neutron star mergers (Hansen \& Lyutikov 2001), primordial black holes (Rees 1977; Blandford 1977; Kavic et al. 2008), and flaring stars (Loeb et al. 2013). However a highly likely RFI candidate is a meteor reflection. The ionized trails of meteors have long been known to reflect man-made RFI. In particular there is a population of long-duration meteor reflections which last up to several minutes and have similar temporal evolution to these two events (Bourdillon et al. 2005). These meteor reflections tend to be linearly polarized, a property the 121118 event is lacking and the 121024 displays no more than what we expect from leakage. 

\section{Rate Density Limits}

With 112.6 hours of data analyzed at 37.9 MHz, a 6$\sigma$ RMS sensitivity of $\leq$ 1260 Jy above a zenith angle of 60$^{\circ}$, and $\pi$ sr of observable sky, we estimate an event rate of $\leq$ $7.5\times10^{-3}$ yr$^{-1}$ deg$^{-2}$ for events with pulse energy densities $>1.3\times 10^{-22}$ J m$^{-2}$ Hz$^{-1}$ and pulse widths of 5 s.

The same can be done for events at both 52.0 MHz and 74.0 MHz, which have 29.7 and 59.8 hours of analyzed data and 6$\sigma$ RMS sensitivities $\leq$ 1104 and 1440 Jy. The rate densities for these frequencies are $\leq$ $2.9\times10^{-2}$ yr $^{-1}$ deg$^{-2}$ and $\leq$ $1.4\times10^{-2}$ yr$^{-1}$ deg$^{-2}$ for events with pulse energy densities $>1.1\times 10^{-22}$ and $1.4\times 10^{-22}$ J m$^{-2}$ Hz$^{-1}$ and pulse widths of 5 s. This is similar to the rate densities found in the past by similar experiments in this frequency range. A comparison with Kardashev et al. (1977), Lazio et al. (2010), and Cutchin (2011) is shown in Table 3. 

\begin{deluxetable}{ |c| |c| |c| |c| |c| }
\tablecaption{Pulse Rate Limits}
\tablewidth{0pt}

\tablehead{\colhead{Name} 		&	\colhead{Frequency}	&	\colhead{Rate Density}		&	\colhead{Pulse Energy Density}		&	\colhead{Pulse Width}\\    	&	  \colhead{(MHz)}	&	  \colhead{(yr$^{-1}$ deg$^{-2}$)}	&	  \colhead{(J m$^{-2}$ Hz$^{-1}$)}  }

\startdata
Kardashev et al. (1977)	&	60			&	$\sim10^{-3}$ 			&	$3.1\times 10^{-22}$			&	0.5 s\\
                                      	&	38			&	$\sim1.5\times10^{-3}$	&	$2.1\times 10^{-22}$ 		&	0.5 s\\
Lazio et al. (2010)		&	73.8			&	$\sim10^{-2}$ 			&	$1.5\times 10^{-20}$ 		&	300 s\\
Cutchin (2011)			&	38			&	$\sim2.5\times10^{-1}$	&	$2.6\times 10^{-23}$			&	3 s\\
This Paper			&	37.9			&	$\sim7.5\times10^{-3}$	&	$1.3\times 10^{-22}$			&	10 s\\
                                             	&	52.0			&	$\sim2.9\times10^{-2}$	&	$1.1\times 10^{-22}$			&	10 s\\
					&	74.0			&	$\sim1.4\times10^{-2}$	&	$1.4\times 10^{-22}$ 		&	10 s\\
\enddata

\end{deluxetable}

\section{Discussion}

We have carried out a search for prompt low frequency radio emission from 34 GRBs at 37.9, 52.0, and 74.0 MHz. In this search we found no burst-like emission but have placed limits at these frequencies. Our $1\sigma$ limits for each frequency are listed in Table 1 and range from $\sim$ 200-80 Jy, for $\geq$ 5 second bursts. The range of $DM$s that we are sensitive to depends on the duration of the burst. For 5 second bursts we could see to a maximum of 220, 570, and 1,600 pc cm$^{-3}$ for 37.9, 52.0 and 74.0 MHz.

While these limits do not disprove any of the possible emission mechanisms discussed in the introduction of this paper these are the most stringent to this date. In the future we plan to improve our sensitivity by applying deconvolution and phase calibration to our images.

We also report two transient events, 121024 and 121118, at 37.9 and 29.9 MHz respectively, that lasted for 75 and 100 seconds. We limit their $DM$s to be approximately $\leq$ 450 and $\leq$  250 pc cm$^{-3}$. 

We also have placed rate density limits on general transients with pulse energy densities $>1.3\times 10^{-22}$,  $>1.1\times 10^{-22}$, and $1.4\times 10^{-22}$ J m$^{-2}$ Hz$^{-1}$ and pulse widths of 5 s at 37.9, 52.0, and 74.0 MHz. Using the entire sky higher than $30^{\circ}$ above the horizon we find a maximum rate limits of $\leq$ $7.5\times10^{-3}$, $2.9\times10^{-2}$, and $1.4\times10^{-2}$  yr$^{-1}$ deg$^{-2}$ for the frequencies above.

If it is true that we should see one FRED transient for every $\leq$ 115 hours of observation at 37.9 MHz then a full analysis on the 1000s of hours of data PASI has collected at this frequency should yield several more. A forthcoming paper will address the results of such a large scale search.

\section{Acknowledgements}

Construction of the LWA1 has been supported by the Office of Naval Research under Contract N00014-07-C-0147. Support for operations and continuing development of the LWA1 is provided by the National Science Foundation under grants AST-1139963 and AST-1139974 of the University Radio Observatory program. 

This research has made use of the NASA/IPAC Extragalactic Database (NED) which is operated by the Jet Propulsion Laboratory, California Institute of Technology, under contract with the National Aeronautics and Space Administration.

Part of this research was conducted at the Jet Propulsion Laboratory, California Institute of Technology, under contract to NASA

\end{document}